**Title**

Conflation of short identity-by-descent segments bias their inferred length distribution

**Authors and Affiliations**

Charleston W.K. Chiang,[1] Peter Ralph,[2] John Novembre,[3]

[1] Department of Ecology and Evolutionary Biology, University of California, Los Angeles, Los Angeles, CA 90095, USA.

[2] Department of Molecular and Computational Biology, University of Southern California, Los Angeles, CA 90089, USA

[3] Department of Human Genetics, University of Chicago, Chicago, IL 60637, USA

**Correspondence:** chiang82@ucla.edu, jnovembre@uchicago.edu

**Abssegment**

Identity-by-descent (IBD) is a fundamental concept in genetics with many applications. In a common definition, two haplotypes are said to contain an IBD segment if they share a segment that is inherited from a recent shared common ancestor without intervening recombination. Long IBD segments (> 1cM) can be efficiently detected by a number of algorithms using high-density SNP array data from a population sample. However, these approaches detect IBD based on contiguous segments of identity-by-state, and such segments may exist due to the conflation of smaller, nearby IBD segments. We quantified this effect using coalescent simulations, finding that nearly 40% of inferred segments 1-2cM long are results of conflations of two or more shorter segments, under demographic scenarios typical for modern humans. This biases the inferred IBD segment length distribution, and so can affect downstream inferences. We observed this conflation effect universally across different IBD detection programs and human demographic histories, and found inference of segments longer than 2cM to be much more reliable (less than 5% conflation rate). As an example of how this can negatively affect downstream analyses, we present and analyze a novel estimator of the de novo mutation rate using IBD segments, and demonstrate that the biased length distribution of the IBD segments due to conflation can lead to inflated estimates if the conflation is not modeled. Understanding the conflation effect in detail will make its correction in future methods more tractable.

2**Abssegment**

Identity-by-descent (IBD) is a fundamental concept in genetics with many applications. In a common definition, two haplotypes are said to contain an IBD segment if they share a segment that is inherited from a recent shared common ancestor without intervening recombination. Long IBD segments (> 1cM) can be efficiently detected by a number of algorithms using high-density SNP array data from a population sample. However, these approaches detect IBD based on contiguous segments of identity-by-state, and such segments may exist due to the conflation of smaller, nearby IBD segments. We quantified this effect using coalescent simulations, finding that nearly 40% of inferred segments 1-2cM long are results of conflations of two or more shorter segments, under demographic scenarios typical for modern humans. This biases the inferred IBD segment length distribution, and so can affect downstream inferences. We observed this conflation effect universally across different IBD detection programs and human demographic histories, and found inference of segments longer than 2cM to be much more reliable (less than 5% conflation rate). As an example of how this can negatively affect downstream analyses, we present and analyze a novel estimator of the de novo mutation rate using IBD segments, and demonstrate that the biased length distribution of the IBD segments due to conflation can lead to inflated estimates if the conflation is not modeled. Understanding the conflation effect in detail will make its correction in future methods more tractable.

**Introduction**

A genomic region is said to be identical-by-descent (IBD) between a pair of individuals if the region was co-inherited from a common ancestor without any intervening recombination. This and related concepts of co-inheritance and IBD have been useful in population genetic theory, as evident by their numerous genetic applications. For example, identifying shared haplotypes forms the basis of several imputation (GUSEV *et al.* 2012) and phasing (KONG *et al.* 2008; BROWNING AND BROWNING 2011b) methods; the frequency of IBD within and between cohorts allows the detection of natural selection and trait-associated loci (PURCELL *et al.* 2007; ALBRECHTSEN *et al.* 2010; GUSEV *et al.* 2011; HAN AND ABNEY 2013); contrasting the number and the length of IBD segments enables inferences of past demographic histories (PALAMARA *et al.* 2012; RALPH AND COOP 2013). The concept of IBD has also been used to estimate key human genetic parameters such as the mutation rate (CAMPBELL *et al.* 2012) and heritability (ZUK *et al.* 2012).

Traditionally the concept of IBD regions or segments has been defined with respect to a set of founder individuals, such as in a pedigree (THOMPSON 2013). More recently, attention has been paid to IBD segments among pairs of individuals in populations without known pedigree. In finite populations all gene copies trace their ancestry to a single common ancestor, such that any individual is effectively IBD with everyone else throughout the genome, sharing a common ancestor at some point in the past. Most of the IBD segments arising in random samples are short relative to the spacing of polymorphic sites (CHAPMAN AND THOMPSON 2003; POWELL *et al.* 2010; THOMPSON 2013), having been broken up by many recombination events over time. Therefore, in practice, to avoid studying segments of similarity that might arise due to chance, analyses typically identify long IBD segments, *i.e.*, the subset of all putative IBD segments that exceed a particular length threshold. A typical length threshold used is 1cM (BROWNING AND BROWNING 2011a). However, this threshold for detection is arbitrary, and depends on the behavior of the specific algorithm one uses and the research question one seeks to answer.

A number of computer programs exist to detect IBD segments using high-density array data from population samples. Beagle IBD (BROWNING AND BROWNING 2010), RELATE (ALBRECHTSEN *et al.* 2009), IBDLD (HAN AND ABNEY 2011), and PLINK (PURCELL *et al.* 2007) detect IBD segments based on a probabilistic hidden Markov model for IBD status to determine the posterior probability of IBD at a genomic location. Probabilistic models tend to be computationally intensive (BROWNING AND BROWNING 2013b), and thus often are infeasible to apply to large population datasets. On the other hand, programs like GERMLINE (GUSEV *et al.* 2009), Beagle fastIBD (BROWNING AND BROWNING 2011a) and Refined IBD (BROWNING AND BROWNING 2013b) adopt a dictionary approach to detect IBD, followed by varying degrees of probabilistic assessment of IBD segments to improve accuracies. These dictionary approaches are scalable to large sample datasets. Note that in all cases these methods identify IBD segments by looking for long stretches of near-identical sequences (identity-by-state, or IBS). Some applications then assume that such segments, if long enough, have been uninterrupted by intervening recombination (and thus co-inherited from a single common ancestor).

This assumption, as we will examine in the current manuscript, does not always hold. It is clear that long segments of IBS could be due to the conflation of two or more shorter IBD segments, especially using diploid data, where the two IBD segments could be shared between different haplotypes. This error in the estimated length will in turn lead to errors in the estimates ages of common ancestors and bias downstream inferences. For example in demographic inference, the conflation could lead to an inference of more recent coalescent events, and hence a smaller effective population size, than in fact existed if this is not explicitly modeled. Similarly, and as we will show, an approach to estimate the mutation rate using the length of IBD segments would erroneously assign younger ages to conflations of older segments and thus overestimate the mutation rate.

We used coalescent simulations where we know the precise ancestral recombination graph (ARG (HUDSON 1991; GRIFFITHS AND MARJORAM 1996)) of the simulated sample to test our hypothesis. At all positions of a simulated sequence, the ARG relates individual haplotypes via a local genealogical tree. An IBD segment shared between two individuals in the simulation can be identified as all the stretch of contiguous sequence where the subset of the ARG relating two haplotypes from the two individuals does not change (**Figure 1**). This approach of identifying ground-truth IBD segments is very similar to a framework recently developed to test the power and accuracy of IBD detection programs (BROWNING AND BROWNING 2013b).

We then simulated genotype data, from which we inferred IBD segments using published algorithms, and contrast the algorithm-detected IBD segments to the segments with recent common ancestry implied by the underlying simulated ARGs. We show that under various demographic models appropriate for modern humans, >35% of the detected IBD segments of 1cM or longer (predominately due to those between 1-2 cM long) are composed of at least two subsegments. As an example of how the conflation can lead to practical problems, we analyze the behavior of a potential estimator of the *de novo* mutation rate using IBD segments. We observed accurate estimates of the input mutation rate when true IBD segments are used, but overestimates of the mutation rate by ~10-40% using inferred IBD segments, depending on the amount of conflation in the dataset. Note that the main algorithm of IBD detection tested here is Beagle's fastIBD (BROWNING AND BROWNING 2011a), though we nevertheless tested a number of other algorithms designed to detect IBD segments from array data, and demonstrate that the conflation problem is faced by all algorithms tested.

**Material and Methods**

*Coalescent simulations*

We simulated five 20Mb regions of 2000 haplotypes each using MaCS (CHEN *et al.* 2009). To mimic genetic data of actual populations, we assumed in our simulation a demographic model previously estimated for a European population (NELSON *et al.* 2012). Specifically, we modeled a population with constant size 12,500 up until 17,000 generations ago, 24,500 between 3,500 and 17,000 generations ago, 7,700 between 368 and 3,500 generations ago, and exponentially expanding at a rate of 0.017 per generation to a present day size of $4 \times 10^6$. The MaCS command used is given in the **Appendix**.

To generate phased sequence data from the MaCs output, we paired the genotypes from two randomly chosen simulated haplotypes. We introduced no missing data or genotyping errors and assume a constant recombination rate. To generate genome-wide array data, we down-sampled the sequencing data to match the marker density and allele frequency spectrum typical of array data. For our parameter settings, MaCS produced between 254K to 256K variants per simulated region (**Table S1**). The marker density was 5,800 markers per 20 Mb region, which extrapolates to approximately 715K markers across the entire set of autosomes for which recombination rate estimates are available (approximately 2,450 Mb). This density is typical of a genome-wide 1M array after quality controls and filtering on common variants. The minor allele frequency spectrum was empirically matched to that found on an array, based on the array data from European individuals (i.e. approximately a flat distribution for variants > 5% frequency). Markers with minor allele frequency < 5% were dropped from the array data. In total, 5.7K to 5.9K variants were retained for the array data for each simulated region (**Table S1**). We used simulated array data to call IBD segments with a number of available IBD calling algorithms, and then used the simulated sequencing data to infer mutation rate (below).

In addition to the plausible demographic model from Nelson et al. (NELSON *et al.* 2012), we also tested a number of other demographic trajectories: the European and African demography from Tennessen et al. (TENNESSEN *et al.* 2012), a European model but without the aggressive second exponential growth phase, as well as a constant size population. Detailed descriptions of the demography and MaCS commands used for these scenarios can be found in the **Appendix**.

*Detecting IBD$_{called}$ segments*

We combined simulated haploid genotypes at random into pairs to produce unphased diploid individuals. We then use available IBD detection programs to look for IBD segments between pairs of diploid individuals.

The main algorithm of IBD detection tested here is Beagle's fastIBD (BROWNING AND BROWNING 2011a), though we nevertheless tested a number of other algorithms designed to detect IBD segments from array data, namely Refined IBD (BROWNING AND BROWNING 2013b),

GERMLINE (GUSEV *et al.* 2009), IBDLD (HAN AND ABNEY 2011), and PLINK (PURCELL *et al.* 2007). Even though the fastIBD algorithm has since been superseded by Refined IBD, it remains a popular algorithm used in large-scale genetic mapping studies due to its speed and scalability to large datasets. Furthermore, Refined IBD does not allow for genotyping errors, which may not be realistic in high-density array datasets.

As recommended by the authors of fastIBD/Beagle, IBD estimations for each simulated region was conducted ten independent times and then combined (BROWNING AND BROWNING 2011a), retaining only segments supported by at least two independent runs and at least one segment reaching a confidence score below $10^{-10}$. Others (BROWNING AND BROWNING 2013b; RALPH AND COOP 2013; DURAND *et al.* 2014) have noticed a tendency of spurious gaps or breaks introduced into long IBD segments by Beagle due to low marker density, switch errors, or genotyping errors and have then chosen to merge nearby segments separated by a gap shorter than at least one of the two segments. In our evaluation, we found that much of the need for merging may have come from genotyping errors or missingness (data not shown). As we assumed perfect genotyping data, such merging could artificially induce the phenomenon of interest here, namely the conflation of two segments. As such, we elected to be more conservative and simply removed approximately 1.3% of segments that would otherwise be merged by the criteria presented in (RALPH AND COOP 2013) from our analyses. To avoid artifacts due to the edges, we also removed segments that overlapped the boundary of the simulated region. In total, we were left with 3,003 IBD segments greater than 1 cM in length across the five simulated regions (**Table S1**).

To compare to IBD segments called by other algorithms, we also used GERMLINE (v.1.5.1), Refined IBD (v.1128) and IBDLD (v.3.004.4) to call IBD segments (GUSEV et al. 2009; BROWNING AND BROWNING 2013b; HAN AND ABNEY 2013). We applied each method to a single simulated region to facilitate comparisons across methods in a computationally feasible manner.

To estimate IBD segments using GERMLINE, we first subjected the simulated data to 10 iterations of phasing by Beagle, before processing through GERMLINE. Default parameters were used, except that we required the minimum length of the segment to be 1 Mb (min_m = 1), and that the haploid mode (--haploid) was used. The haploid option allows GERMLINE to utilize the haplotype phase information, and should improve performance (BROWNING AND BROWNING 2013b). In total, we detected 99 IBD segments greater than 1cM from one simulated region.

Default settings were used to estimate IBD segments using Refined IBD. IBD detection was run only one time, as the authors find that a single run of Refined IBD achieves comparable if not greater power than 10 runs of fastIBD (BROWNING AND BROWNING 2013b). A total of 1,005 IBD segments greater than 1cM from one simulated region were detected.

We also detected IBD segments using IBDLD. Our initial run using all 1000 diploid individuals for one simulated region did not complete within 72 hrs. Therefore we randomly selected a subset of 500 diploid individuals for our evaluation. We used the method GIBDLD (--method GIBDLD) and chose to output only segments with length > 1000 kb (--length 1000); otherwise, default settings were used. A total of 643 IBD segments were detected.

We had also tested PLINK (v1.07, (Purcell *et al.* 2007)). To detect IBD segments, we first thinned the dataset so that no pair of SNPs within a window of 100 markers (moving at windows of 25 markers) would have an $r^2 > 0.2$, as recommended by the author of PLINK. Estimation of IBD segments was then performed on the thinned dataset, using default settings, but detected no segments. Less stringent thinning, increased marker density, lowered minimum number of SNPs or minimum segment length did not significantly improve detection of IBD segments using PLINK (data not shown). This is consistent with the lower power to detect IBD segments by PLINK as observed by others (Browning and Browning 2010), and PLINK was thus not further evaluated.

*Calling true IBD subsegments among $IBD_{called}$ segments*

To determine if an $IBD_{called}$ segment detected by one of the above algorithms is actually composed of two or more true IBD subsegments, we used the following heuristics. As boundaries of called IBD segments tend to have lowered confidence, we first trimmed 10% of the segments from each end (leaving only 80% of the segment length). We then examined all local ARGs output by MaCS within each detected segment. Neighboring ARGs that differ represent a recombination event that altered the topology of successive graphs, though the recombination event need not involve any of the four haplotypes composing the two putative diploid IBD individuals. For each of the four pairwise haplotype configurations between the two putative IBD individuals we estimated the age of common ancestor given the branch length of each of the local genealogy. An $IBD_{ARG}$ (sub-)segment is then defined as continuous stretches of genealogies with unchanging age to common ancestor between at least one pair the four pairwise haplotype configurations. To map each IBD$_{called}$ segments as a composite of IBD$_{ARG}$ subsegments, we took a greedy approach and iteratively include the subsegment with the lowest ratio of age to length from the region until the entire region is exhausted. We validate our heuristic using a set of ground-truth IBD$_{ARG}$ segments > 1cM in length (see below). As expected, each of the IBD$_{ARG}$ segment is composed of one continuous segment, rather than composed of two or more subsegments (data not shown).

*Calling ground-truth IBD$_{ARG}$ segments from simulated data*

We also sought to compare inferred IBD segments with the true IBD segments that exist in the simulated ARGs. It is computationally challenging to enumerate all pairwise IBD segments in an ARG. Therefore, we implemented two approaches to make the problem computationally tractable. The first approach focuses on identifying only the IBD$_{ARG}$ segments that are at most 300 generations old, thereby drastically reducing the number of branches on each genealogy needed to be computed and stored in memory. The second approach samples a genealogy every 0.01 cM (or 10 kb) of the simulated dataset, greatly reducing the number of genealogies that need to be processed. In this case we were able to extract all IBD$_{ARG}$ segments less than 3,000 generations old, which allowed us to obtain all segments with length 0.2 cM or greater (max age = 2,757 generations, compared to segments with length 0.1 cM, which has a max age of 2,994 generations, **Table S2**). However, in this case each IBD$_{ARG}$ segment detected would have an uncertain boundary resolution within 0.01 cM. For comparisons with IBD$_{called}$ segments and for testing the mutation rate estimator framework, the IBD$_{ARG}$ segments < 300 generations old but with precise boundaries were used. For investigating the rate of conflation within our simulated datasets, IBD$_{ARG}$ segments < 3,000 generations old were used.

In general, to extract IBD$_{ARG}$ segments we start from the beginning of the simulated region and for each genealogy we first remove all nodes with age older than 300 or 3,000 generations. Of the remaining nodes, we record the age of all pairwise comparisons of descendants from the node. We then look for continuous stretches of genealogy where a pair of haplotypes showed unchanging age to common ancestor, removing stretches that began or end at the boundary of the simulated region, or those with length below the threshold set by the analysis. While our approach may not detect all IBD$_{ARG}$ segments in the simulated dataset, we can ensure that all detected IBD$_{ARG}$ segments in this manner will be true IBD segments free of any conflation effects.

*Estimating mutation rate based on IBD segments*

We demonstrate the impact of conflating IBD segments on downstream analyses using an IBD-based estimator of the mutation rate. There has been recent interest to estimate mutation rates based on principles of IBD sharing to complement existing approaches from trios and phylogenetic analyses (CAMPBELL *et al.* 2012; PALAMARA *et al.* 2015). Here, our focus is in demonstrating how conflation may bias these estimates; as such, we introduce a simplistic estimator of mutation rate based on IBD, assuming that the phase and demographic history of the sample are known and that no genotyping or sequencing errors exist. Specifically, we imagine a scenario where genome-wide array data are used to identify IBD segments, and then targeted sequencing (e.g. exome sequence data) reveals mutations that have arisen since the common ancestry of the two IBD haplotypes. These mutations must have occurred more recently than the time of the common ancestor giving rise to the IBD. Therefore, an estimator of the mutation rate μ is:

$$\mu = \frac{\sum_{i \in I} m(i)}{\sum_{i \in I} 2L_{seq}(i)T_{IBD}(i)}$$

Eqn.1

where *I* is the set of all IBD segment, and for a given IBD segment *i*, *m(i)* is the number of sequence mismatches on the two IBD haplotypes, $L_{seq}(i)$ is the total sequenced region (= the length of the IBD segment, $L_{IBD}(i)$, if completely sequenced), and $T_{IBD}(i)$ is time since the common ancestor in generation. In practice, $T_{IBD}$ is rarely known for the sample, and estimation of the time to the most recent common ancestor can be difficult. For our purpose here to illustrate the impact due to conflation, we estimate $T_{IBD}$ based on simulations using a fixed demographic model (below).

*Simulating conflated segments to test the impact of conflation on mutation rate estimator*

Based on all of the $IBD_{ARG}$ segments with length > 0.2 cM, we calculated the mean and standard deviation of ages in bins of 0.1 cM (**Table S2**). This is the age distribution as a function of IBD segment length. In practice, this distribution will change depending on the underlying demography of the population. However, for the purpose of our illustration we assumed that the demography is known.

Then, to simulate conflated segments, we sampled IBD segment lengths from the apparent length distribution as well as with the proportion of conflation we estimate from our simulations (see **Results**). We assigned each sampled IBD segment as a conflated segment or a non-conflated segment, depending on its length and the appropriate proportion of conflation given its length. For non-conflated segments, we sampled an age, $T_{UC}$, from the age distribution given its length, $L_{UC}$, assuming the age is normally distributed with means and standard deviation shown in **Table S2**. To simulate the number of sequence mismatches that would be observed in this IBD segment, we drew random numbers from a Poisson distribution with mean $2L_{UC}T_{UC}\mu$, where $\mu$ is the mutation rate and set to $1.2 \times 10^{-8}$.

For conflated segments, we make the simplifying assumption that the entire segment of length $L_C$ is composed of only two subsegments of length $L_{C1}$ and $L_{C2}$, with no gap. $L_{C1}$ is sampled from the true length distribution of IBD$_{ARG}$ segments, while $L_{C2} = L_C - L_{C1}$. For each subsegments, we then sampled ages $T_{C1}$ and $T_{C2}$, and assigned number of sequence mismatches according to the same Poisson distribution above.

For both conflated and non-conflated segments, we also sampled the apparent age based on the apparent length (the full length for non-conflated segments, and the sum of the two lengths for conflated segments). These will be the estimated $T_{IBD}$ used in eqn1.

Note that our strategy here is aimed to examine the potential impact of conflation, and conflation alone, on our mutation rate estimator in a conservative manner. As such, we assumed that there are no genotyping or sequencing errors, the phase and the demography are known, and there are no mis-estimation of the IBD segment boundaries and no gaps between the conflated segments. In practice, these are all issues that can further confound the analysis.

**Results**

*The prevalence of subsegment conflations among called IBD segments*

Across the simulations, fastIBD called a total of 3,003 IBD$_{called}$ segments (**Table S1**). To estimate the proportion of these IBD$_{called}$ segments that are composed of multiple subsegments, we inferred the number of subsegments uninterrupted by recombination in the history of our simulated sample (**Material and Methods**). We observed a number of IBD$_{called}$ segments being composed of numerous subsegments (A random subset of 250 IBD$_{called}$ segments is graphically displayed in **Figure 2**; the entire dataset is displayed in **Figure S1**). Because statistical inferences of IBD is most unreliable at segment boundaries (BROWNING AND BROWNING 2013b), we discarded 10% of the segment length from both ends and still found ~37% of the IBD$_{called}$ segments being composed of two or more subsegments (**Figure 3A**). Most importantly, of the IBD$_{called}$ segments consisting at least two subsegments, the second longest subsegment is at least 1/5 of the total called length 35% of the time (13% of all segments), or at least 1/3 of the total called length 12% of the time (4% of all segments) (**Figure 3B**). This suggests that the subsegments are often quite long and represent stretches of IBD inherited from a relatively recent, though different, common ancestor, rather than just statistical noise or inaccuracy in IBD detection. One illustration of a conflated segment is shown in **Figure 4**, where between the four possible pairs of haplotypes between two diploid individuals, two separate segments were inherited from two sets of recent, but different, common ancestors that are in close proximity of each other, leading the algorithm to call the entire region IBD. The conflation effect is also more pronounced among shorter IBD$_{called}$ segments (those between 1cM and 2cM), but is still detectable for segments greater than 2cM (**Figure 3A**).

While we have thus far focused on IBD$_{called}$ segments produced by Beagle's fastIBD program, our observation should not be specific to Beagle, but is more generally applicable to any IBD calling algorithm so long that these programs depend solely on sequence identity for inference. To test this hypothesis, we also used Refined IBD (BROWNING AND BROWNING 2013b), GERMLINE (GUSEV *et al.* 2009), and IBDLD (HAN AND ABNEY 2013) to call IBD segments in each of our simulated regions. We observed the same conflation effects across these algorithms (**Figure S2**, **Figure S3**), though different algorithms were affected to different degrees.

Finally, the conflation effect is also not specific to the demography we simulated. In addition to the European-like human demography estimated in Nelson et al. (NELSON *et al.* 2012), we also simulated under both the European and African demographic histories published in Tennessen et al. (TENNESSEN *et al.* 2012), an alternative European-like model without the most recent exponential growth phase, as well as a constant size population. Though different demographic scenarios are affected by the conflation events to different degrees, the conflation effect is evident in all cases (**Figure S4**). Notably, in scenarios with recent rapid growth a larger fraction of segments are found to be due to conflation of smaller segments.

*The rate of conflation of true IBD segments in the simulated ancestral recombination graphs*

Though we have demonstrated that conflations of significantly lengthy subsegments are prevalent among IBD segments called by a number of algorithms, it remains possible this is particular to the technical aspects of the calling algorithm, rather than a general, genetic, phenomenon. Thus, we next aim to characterize the rate of conflation events in the ancestral recombination graphs using our simulations, removing the imprecision of the calling algorithms, and evaluate the impact of the conflations on the length distribution of IBD segments. We define as the occurrence of a conflation event in our simulated data when two $IBD_{ARG}$ segments, each of length at least $w$ cM, separated by a gap no greater than a small number $g$ cM, together account for an end-to-end length of at least 1 cM. We then computed the rate of conflation from one of our five replicate 20Mb simulated regions. Furthermore, the simplest model of IBD segment conflation is that, given the length distribution determined by the coalescent time distribution, observed long IBD segments are uniformly and independently distributed along the genome. To test this, we simulated 100 replicate datasets by sampling from the length distribution and assigning these randomly (with replacement) to two diploid individuals uniformly across the genomic regions.

We observe that the conflation rate is noticeably higher with the inclusion of increasingly smaller $IBD_{ARG}$ segments (**Figure 5A**, **Figure S5**). When counting conflation events due to segments as small as 0.2 cM, the rate of conflation (0.547 events per 1000 pairs of individuals) is actually on the same order of magnitude as the true IBD rate > 1 cM in length (1.93 events per 1000 pairs of individuals) in the simulated dataset. When comparing to the simulated replicates, the levels of conflation are comparable (**Figure 5A**), suggesting that the majority of the observed conflations are due to the random independent placement of IBD segments along chromosomes. Therefore, any potential correlation in long segment placement induced by the haploid coalescent or the pedigree structure appears minimal.

To further investigate this, one can also assign each conflated segment as being in cis or trans based on whether the conflated pairs of segments involved 2, 3, or 4 of the possible four haplotypes: involving the same two of the four haplotypes implies a *cis* pair in which the same two haplotypes descend from recent but different common ancestors in each segment; while if the pairs involve three or four of the possible haplotypes this suggest the conflated IBD events are in *trans* (**Figure S6**). We find that approximately 80% of the conflations occur *in trans* rather than *in cis* (0.436 events vs. 0.110 events per 1000 pairs of individuals); the proportion predicted by independent placement, 75%, is within the confidence interval (there are three times as many possible *trans* arrangements as *cis*).

The observed rate of conflation leads to a noticeable bias in the length distribution of apparent IBD segments with length greater than 1cM (**Figure 5B**, **Figure S5**). Specifically, approximately 27% of the apparent IBD segments between 1 to 1.2 cM long are due to conflation.

*Evaluating the impact of conflation on a mutation rate estimator*

We evaluated the impact of the biased length distribution of IBD segments due to conflation on a mutation rate estimator. Given accurately defined IBD segments and perfect sequence and phase information, the sequence mismatches between a pair of individuals in the IBD segment reveals mutations that have arisen since the common ancestry of the two IBD haplotypes. Therefore, we devised a simplistic mutation rate estimator based on observed sequence mismatches between a pair of IBD haplotypes (**Methods**). As we are only interested in examining the potential impact of conflation on this estimator, we conservatively simulated IBD segments assuming that when conflation occurs, only two subsegments (whose relative phase is known) compose the observed IBD segment, with no gap in between (**Methods**). We then applied our mutation rate estimator based on the observed sequence mismatches (again known with no error), and the apparent age of the segment (sampled from the same age distribution). The proportion of our simulated segments that are conflated depends on its apparent length, and follows the proportion we estimated in our simulation (**Figure 5B**).

If we applied the mutation rate estimator on a set of non-conflated segments (*i.e.* all IBD segments are called perfectly), we would estimate the same mutation rate ($1.199 \times 10^{-8}$/bp/gen; 5th and 95th percentiles = $1.169 \times 10^{-8}$ and $1.230 \times 10^{-8}$ /bp/gen, respectively) as the value used to simulate the data ($1.20 \times 10^{-8}$ /bp/gen). If the estimator is applied on a set of conflated segments, all composed of two subsegments, we would estimate μ to be $1.652 \times 10^{-8}$ /bp/gen (5th and 95th percentiles = $1.597 \times 10^{-8}$ and $1.723 \times 10^{-8}$ /bp/gen, respectively), which is a 38% inflation. If we applied the estimator to a mixture set composed of ~77% unconflated segments and ~23% conflated segments (as would be predicted by **Figure 5B**), we calculate μ to be $1.306 \times 10^{-8}$ /bp/gen (5th and 95th percentiles = $1.275 \times 10^{-8}$ and $1.330 \times 10^{-8}$/bp/gen), a more modest 9% inflation.

**Discussion**

We have shown here through simulations that irrespective of the choice of IBD-calling algorithm examined, unless one restricts to IBD segments >2 cM, an appreciable proportion of called IBD segments are composed of conflations of shorter, more ancient, subsegments. This is a major issue for approaches that assume such segments are inherited from single common ancestors. For instance, of segments inferred using Beagle's fastIBD greater than 35% are actually composed of two shorter subsegments with intervening regions of very short segments. The rate of conflation is reduced to 5% for longer IBD segments (> 2cM) (**Figure 3**). In our simulations, this resulted from chance conflations between independent, shorter segments (e.g. two of sizes 0.5cM); this creates more segments near the shorter end of our detection threshold (1cM) than longer because the true segment length distribution is so strongly peaked near zero (**Figure 5**). In real data, other factors, such as marker density, could also be in play.

This observation underscores how important it is for methods using information in the IBD segment length distribution to explicitly take into account detection power and false positive rates. A first step is to restrict analysis to segments longer than 2cM, rather than 1cM, as at least in demographic scenarios similar to western Europe, the conflation rates are fairly small. A more careful approach is to explicitly estimate power and false positive rates as a function of segment length (RALPH AND COOP 2013).

The conflation effect described here could also at least partly explain the recent observation of pervasive false-positive IBD segments detected in pedigrees (DURAND *et al.* 2014). In Durand et al., the authors noted that over 67% of the IBD segments they identified with GERMLINE between a child in a trio to an unrelated individual are not observed between either of the parent in the trio to the same unrelated individual. This false positive effect is also exacerbated near the algorithm's length threshold for making a call. Conceivably, at least a subset of these false positive events is not due to blatant errors in the calling algorithm. In fact, GERMLINE performed comparably well for the IBD segments it did detect in our simulation (**Figures S2, S3**). Instead, some of the IBD segments called in the child-to-other comparisons could be due to conflations of IBD segments, independently inherited from the two parents. As a result, the full length IBD segments are not observed in either of the parent.

The failure to account for the biased length distribution of IBD segments due to conflation would lead to errors in downstream analysis, which we demonstrated with our estimator of the mutation rate based on IBD segments. The mutation rate is a fundamental parameter in the study of genetics and evolution, on the basis of which much of the population genetic theory and studies of species evolution are built upon (SCALLY AND DURBIN 2012). There are a number of advantages to having an estimator of µ based on IBD segments. Traditionally, estimates of µ have been derived using divergence of neutral sites between species (NACHMAN AND CROWELL 2000; KUMAR AND SUBRAMANIAN 2002). More recently, family-based approaches based on direct sequencing in pedigrees (AWADALLA *et al.* 2010; ROACH *et al.* 2010), or population-based approaches modeling the demographic history of a population and analyzing its impact on the site frequency spectrum (COVENTRY *et al.* 2010; NELSON *et al.* 2012) have also been applied. However, divergence-based estimates rely on accurately knowing the speciation time, family-based estimates are limited to information from only a few families, and population-based estimates are sensitive to the assumed demographic model. Therefore, an IBD-based approach for estimating mutation rates in a population would be useful to complement the current approaches (PALAMARA *et al.* 2015).

While we have shown that conflation is one of the hurdles to overcome in constructing an estimator of the mutation rate based on IBD, other hurdles exist. For example, estimating the age of the MRCA, particularly based on estimated segments, is likely difficult. In the present paper we used simulations to estimate the mean age of the IBD segment as a function of their length. However, our approach depends on knowing, at least approximately, the demographic model. In practice, one rarely knows the demography of the sample, and therefore one approach would be to first infer plausible demographic models of the sample, which can then be used in coalescent simulations to predict the effect of conflations of subsegments on the apparent distribution of ages of the detected IBD segment. Another major hurdle is the relative phase between the two conflated segments. As we observed that the majority of the conflations are due to subsegments *in trans*, a comparison between presumptive IBD haplotypes between two individuals may erroneously include large regions of non-IBD (as the IBD region is now shifted to a different phase). This is potentially problematic if the dataset used to call IBD (*e.g.* genome-wide array data) is phased separately from the sequence mismatch data (*e.g.* exome data of the IBD region). Indeed, these issues and modeling of genotyping errors are the focus of recent attempt to estimate mutation rate using IBD segments (PALAMARA *et al.* 2015). However, Palamara *et al.*'s approach relied on trio-phasing, which limits the sample size available for analysis. Based on our analysis, the potential of conflation may partially explain the slightly elevated estimates in Palamara *et al* (~$1.7 \times 10^{-8}$ /bp/gen) relative to recent pedigree-based estimates (~$1.2 \times 10^{-8}$ /bp/gen).

    Other types of analyses that rely heavily on the age distribution of IBD segments, and thus could potentially be affected by the conflation effect, are demographic inferences based on IBD segments. For example, Ralph and Coop (RALPH AND COOP 2013) devised a non-parametric approach to infer the number and age of recent shared ancestors between populations based on IBD sharing between populations. This provides estimates of distributions of the ages of IBD segments given their lengths, but analytic difficulties inherent to demographic inference may make these quite noisy. Palamara et al. (PALAMARA *et al.* 2012) avoided this by taking a parametric approach, fitting a few parameters to

demographic models based on the age and length distribution of IBD segments. The abundance of recently shared IBD segments provides a window onto the most recent history of a population. However, each of these analyses could be biased if the age distribution of IBD segments were in error due to miscalling of segment lengths. Ralph and Coop focused only on extra long IBD segments > 2 cM, which we show could remove the majority, though not all, of conflation effects. However, this would reduce the available number of IBD segments for analysis, thus reducing power. Instead of using a hard cut-off, Palamara et al. used an approach that iterates between demographic inference and IBD detection until they converged on a set of parameters for GERMLINE that minimized the difference between called and true IBDs from simulation specific to the data at hand, prior to the formal demographic inference procedures. Alternative approaches using IBS directly rather than IBD can also be taken, as done by Harris et al., as the length of IBS segments can be more precisely obtained given high-coverage sequencing data (HARRIS AND NIELSEN 2013). It will be interesting to investigate if a theoretical framework contrasting IBS and apparent IBD segments can predict the rate of conflated segments within a dataset.

Though we focused on an IBD detection framework using high-density array data, as the field develops algorithms aimed to detect even shorter segments due to the improved density of markers in the era of whole genome sequencing (e.g. programs like IBDseq (BROWNING AND BROWNING 2013a)), it seems imperative to note and reevaluate the potential problem of conflation. We discovered that ~80% of the conflation are the result of pairs of segments existing *in trans*, suggesting that improvements in phasing may greatly reduce, but not completely eradicate, biases due to conflated segments. Therefore, in addition to improvements in phasing, considerations of the conflation effect and external information such as availability of other individuals in the pedigree, could all contribute towards improved accuracy in IBD detection and interpretation.

**Appendix**

MaCS commands for simulations

The primary demographic model used throughout the manuscript is one based on deep resequencing of 202 genes across 14,000 European individuals (NELSON *et al.* 2012): the population has ancestral size of 12,500 up until 17,000 generations ago, 24,500 between 3,500 and 17,000 generations ago, 7,700 between 368 and 3,500 generations ago, and exponentially expanding at a rate of 0.017 per generation to a present day size of $4\times10^6$. The MaCS command used is:

```
macs 2000 20000000 -T -t 0.192 -r 0.16 -G 0.017 -eN 0.000023 0.001925 -eN 0.00021875 0.006 -eN 0.0010625 0.003125
```

All input parameters are scaled by $4N_0$ for MaCS, with $N_0$, the present day effective population size, equaling $4\times10^6$ in this case.

Moreover, we simulated under a number of other demographic scenarios, primarily based on those published by Tennessen et al. (TENNESSEN *et al.* 2012), but sometimes with modifications. These include: the European and African demography from Tennessen et al. (TENNESSEN *et al.* 2012), a European model but without the aggressive second exponential growth phase, as well as a constant size population. Specifically, for the European demography we modeled a population with ancestral size 7,310, and has a population size of 14,474 between 2,040 and 5,920 generations ago, 1,861 between 920 and 5,920 generations ago, after which the population experienced exponential growth at rate 0.003 per generation until 204 generations ago, when the growth rate increases to 0.0195 per generation to reach a present day size of 496,500. The African demography was modeled with a population of size 7,310 until 1,480 generations ago, 14,474 between 205 and 1,480 generations ago, and then exponentially growing at rate 0.0165 per generation until reaching a present day size of 424,000. The alternative European model without the aggressive second exponential growth phase is similar, but reaching only a present day size of 9,300. Finally, the constant size population has a constant size of 100,000. The MaCS command for these simulations are:

Tennessen European model ($N_0$ = 496,500):
```
macs 2000 20000000 -T -t 0.023832 -r 0.01986 -G 0.0195 -eG
0.000102718 0.00307 -eN 0.000463746 0.00374823 -eN 0.00102719
0.02915204 -eN 0.002980864 0.01472305
```

Tennessen African model ($N_0$ = 424,000):
```
macs 2000 20000000 -T -t 0.020352 -r 0.01696 -eG 0.0 0.016475022
-eN 0.000120873 0.034136873 -eN 0.003490568 0.017240607
```

Tennessen European alternate model ($N_0$ = 9,300):
```
macs 2000 20000000 -T -t 0.0004464 -r 0.000372 -eN 0.0 1.0 -eG
0.005510753 0.003074847 -eN 0.024758065 0.20010753 -eN
0.05483871 1.55634409 -eN 0.159139785 0.78602151
```

Constant size population model ($N_0$ = 100,000):

```
macs 2000 20000000 -T -t 0.0048 -r 0.004 -eN 0.0 1.0
```

**Description of Supplemental Data**

The Supplemental Data section includes six figures and two tables.

**Acknowledgements**

The authors thank Kirk E. Lohmueller and Alexander Platt for helpful discussions and critical reading of the manuscript. C.W.K.C is supported by an NIH Ruth L. Kirschstein National Research Service Award for postdoctoral fellows (F32GM106656). J.N. is supported by NIH R01 HG007089.

**Figures Titles and Legends**

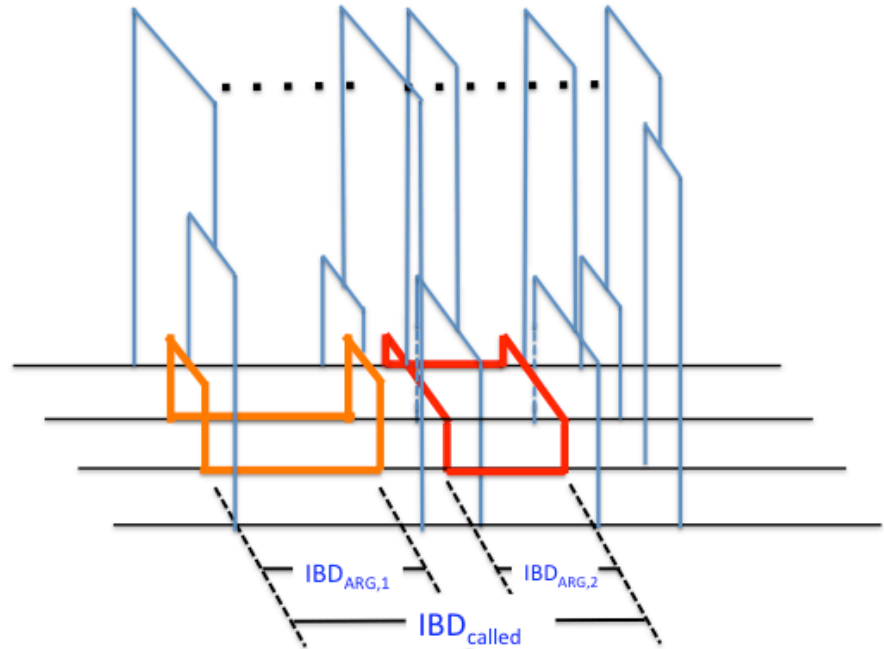

**Figure 1:** a schematic relating IBD$_{ARG}$ and IBD$_{called}$ segments

The cartoon shows the alignment of four haplotypes belonging to two diploid individuals. Across the region multiple ARGs exist to relate the four haplotypes in a tree. Across the first IBD$_{ARG}$ region (orange, IBD$_{ARG,1}$), the two middle haplotypes have recent, unchanging, local graphs for the entire region. Across the second IBD$_{ARG}$ region (red, IBD$_{ARG,2}$), two different haplotypes have recent, unchanging, local graphs for the entire region. The two IBD$_{ARG}$ segments happens to occur near each other such that the entire region may be detected by algorithms based on long stretches of sequence similarity (IBD$_{called}$ segment).

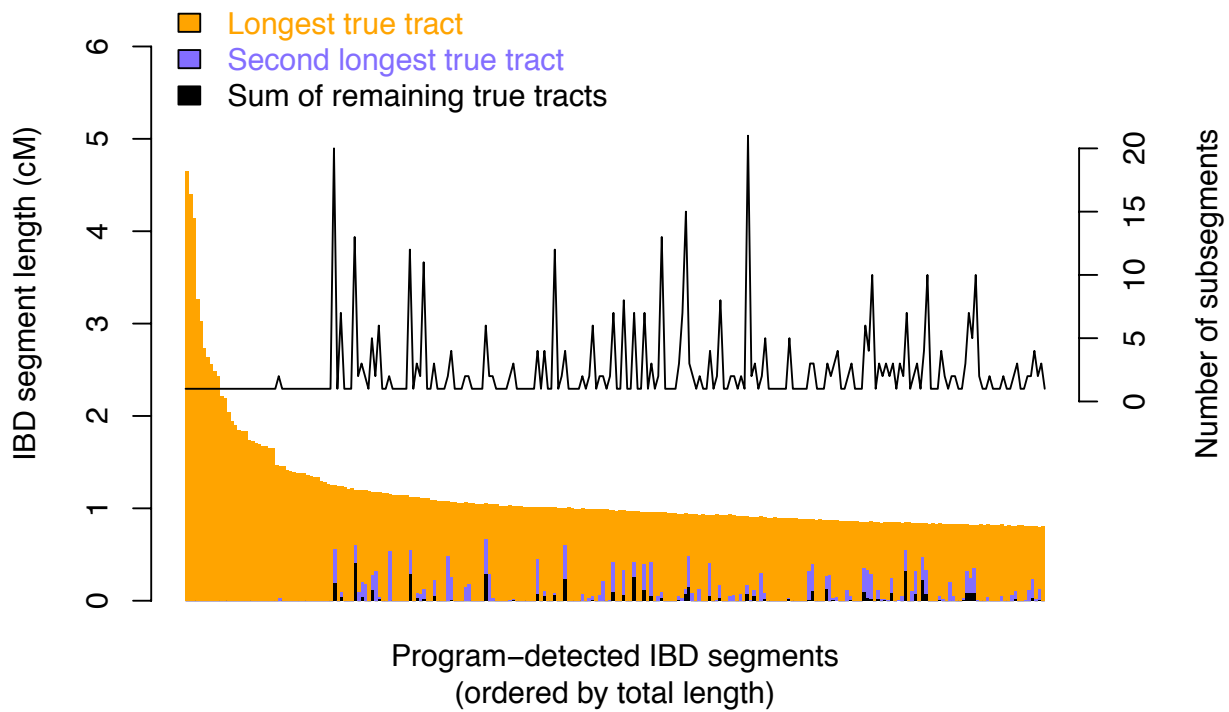

**Figure 2:** The prevalence of subsegments among algorithm-detected IBD segments

Each of a set of randomly chosen 250 IBD$_{called}$ segments detected by Beagle's fastIBD algorithm is represented by a yellow vertical bar. The segments are sorted along the x-axis according to the detected length of the segment in decreasing order. For each segment, the second longest subsegment, if present, is shown in blue, while the remaining subsegments are clumped in black. For each IBD$_{called}$ segment displayed, we also show the number of subsegments detected using the vertical axis on the right. Note that while each IBD$_{called}$ segment was detected by meeting the minimum length threshold of 1cM, we also removed 10% from both ends of the segment for this analysis and thus the shortest segment has a plotted length of 0.8 cM. The choice of trimming 10% of each end is arbitrary, but conservative, as it would decrease our ability to detect conflation events in our dataset. See

**Figures S3** for results based on IBD$_{called}$ segments detected by GERMLINE, Refined IBD, and IBDLD.

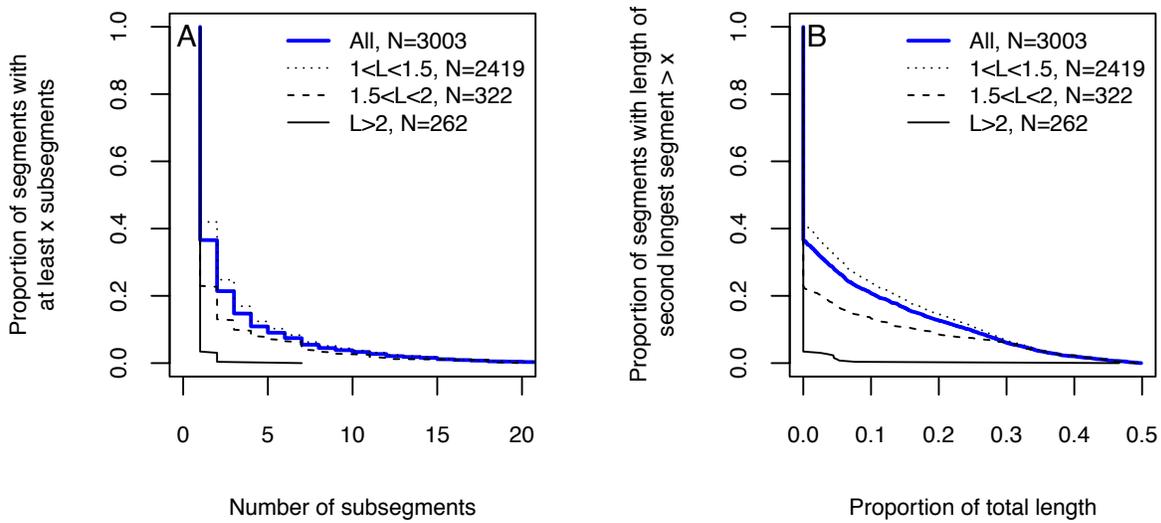

**Figure 3:** The conflation effect is more pronounced for shorter IBD$_{called}$ segments

The complementary cumulative distribution functions (*i.e.,* 1-CDF) for (A) the number of subsegments and (B) the length of the second longest subsegment are shown. The distribution function using all IBD$_{called}$ segments is depicted by a thick blue line. We also show the distributions when IBD$_{called}$ segments are stratified by three levels of length: Between 1 to 1.5cM, dotted line; between 1.5 to 2cM, dashed line; greater than 2cM, solid line. For both measures the conflation effect is more pronounced for the segments between the length of 1 to 1.5cM, though the effect is still detectable among longer segments.

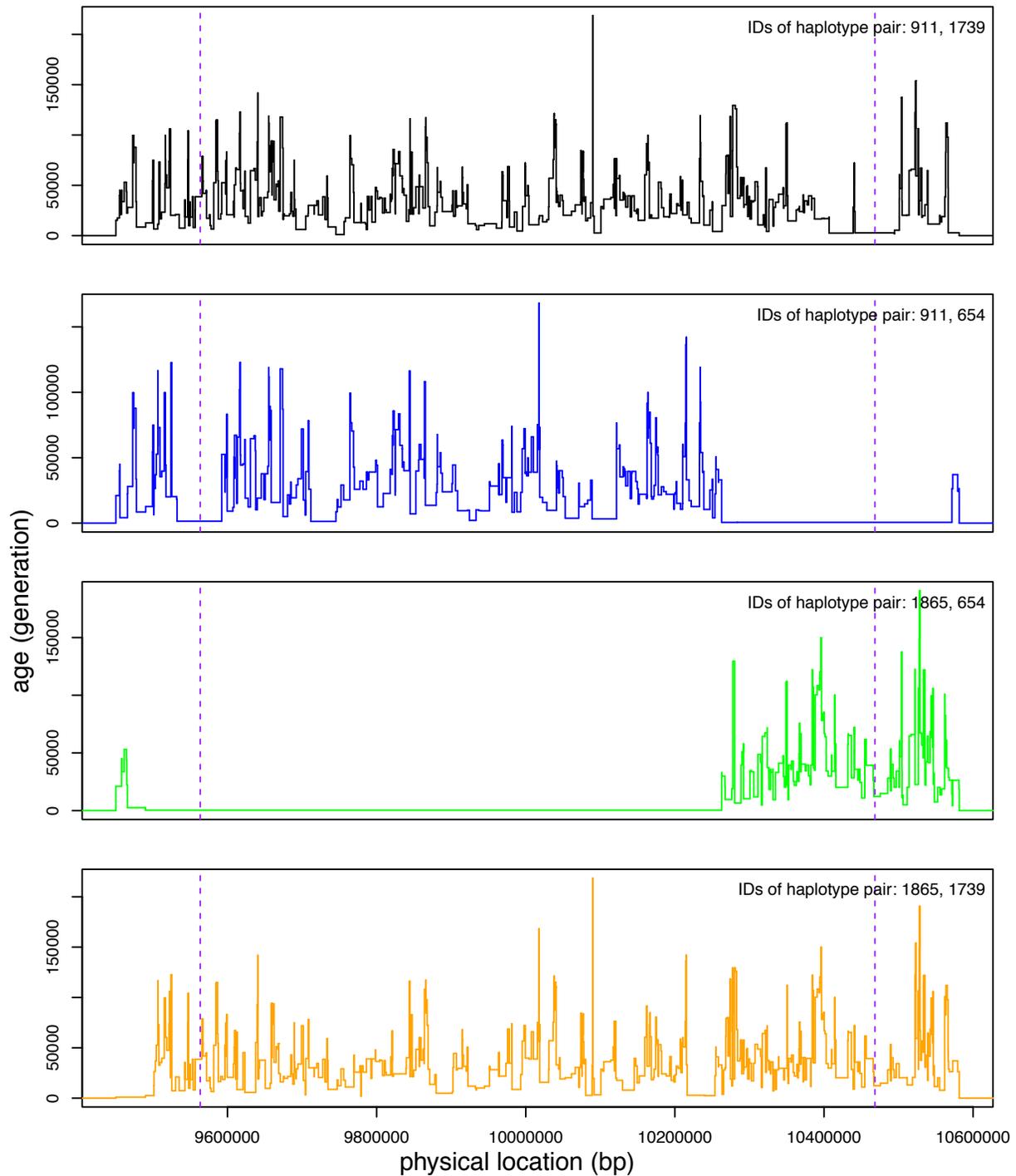

**Figure 4:** An illustrative example of the conflation effect on the mutation rate estimator

For a particular IBD segment of length 1.145 cM detected by Beagle's fastIBD algorithm, we show the distribution of subsegment age (y-axis) as a function of position (x-axis, between 9.45 Mb to 10.59 Mb of a 20-Mb simulated region). Each of the four different pairwise haplotype configurations between the two diploid samples is illustrated with a different color. The simulated haplotype numbers are displayed in the upper right hand corner. The vertical dashed lines demarcate the 10% segment length from both ends of the segment that we would remove from analysis due to the uncertainty in estimating the ends of the $IBD_{called}$ segments. The age of each subsegment is plotted as a step function of its length. In this case, the IBD region is dominated by two long segments of IBD, one between simulated haplotypes 1865 and 654, another between simulated haplotypes 911 and 654. (There is actually a third, very short, segment of recent coalescence between simulated haplotypes 911 and 654 that is not obvious here.) Regions that do not produce long IBD segments can be clearly seen with the deep coalescences. In this case, the predominate IBD haplotype should be between haplotypes 1865 and 654, but the conflation with a neighboring IBD haplotype between haplotypes 911 and 654 led to the estimation of a single long IBD segment.

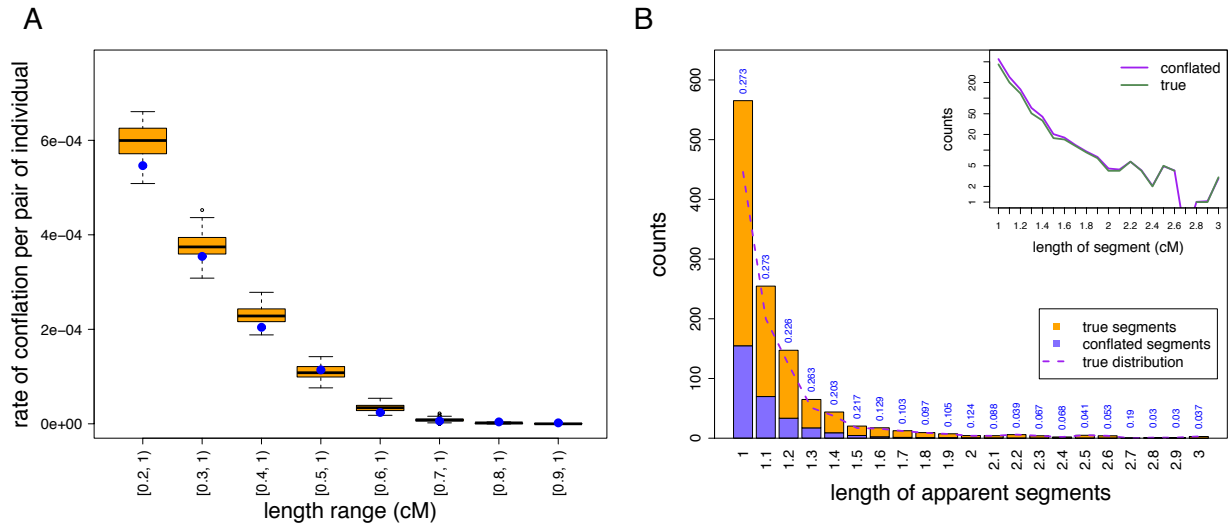

**Figure 5:** Conflations of shorter IBD segments will bias the length distribution

(A) For each bin of segment length range, we calculated the rate at which two IBD$_{ARG}$ segments in our simulations, both within the length range, are adjacent and together constitute an end-to-end length of at least 1cM. The blue dot is the actual value observed in simulation. The boxplot shows the variance around the observed value by randomly sampling from the observed segment length distribution but randomly assigning the location of a segment and sample IDs 100 times. (B) The biased length distribution if each conflated IBD$_{ARG}$ segment is counted for its conflated length rather than the two true lengths. Note that the apparent length of each conflated segment is due to conflation of two IBD$_{ARG}$ segments, independent of any imprecision due to algorithm calling. Dotted line is the true length distribution if each conflated segment can be resolved based on the coalescent genealogy. Inset shows the comparison between the biased length distribution and the true length distribution in log scale. For results based on other parameterizations of the gap sizes (*i.e.*, a maximum gap size of 0.01 cM and 0.05 cM), refer to **Figure S5**.